\documentclass[twocolumn,showpacs,preprintnumbers,prb]{revtex4}
\usepackage{graphicx,bm,amsmath,amssymb}
\usepackage{comment}
\usepackage{hyperref}
\usepackage{xcolor}
\def\gz{\ifmmode{Z\hskip -4.8pt Z}
    \else{\hbox{$Z\hskip -4.8pt Z$}}\fi}

\newcommand{\be}{\begin{equation}}
\newcommand{\ee}{\end{equation}}
\newcommand{\bea}{\begin{eqnarray}}
\newcommand{\eea}{\end{eqnarray}}

\begin{document}

\title{Influence of Rashba spin-orbit coupling on the 0-$\pi$ transition and Kondo temperature in 1D superconductors}

\author{I.~Hamad}
\affiliation{Instituto de F\'{\i}sica Rosario. Facultad de Ciencias Exactas Ingenier\'{\i}a y Agrimensura, Universidad
Nacional de Rosario. Bv. 27 de Febrero 210 bis, 2000 Rosario, Argentina}

\author{F. Lisandrini}
\affiliation{Instituto de F\'{\i}sica Rosario. Facultad de Ciencias Exactas Ingenier\'{\i}a y Agrimensura, Universidad
Nacional de Rosario. Bv. 27 de Febrero
210 bis, 2000 Rosario, Argentina}

\author{C.~Gazza}
\affiliation{Instituto de F\'{\i}sica Rosario. Facultad de Ciencias Exactas Ingenier\'{\i}a y Agrimensura, Universidad
Nacional de Rosario. Bv. 27 de Febrero
210 bis, 2000 Rosario, Argentina}

\author{Alejandro M. Lobos}
\affiliation{Instituto Interdisciplinario de Ciencias B\'{a}sicas, Universidad Nacional de Cuyo, CONICET, Facultad de Ciencias Exactas y Naturales, Padre J. Contreras 1300, (5500) Mendoza, Argentina.}

\begin{abstract}

Using the framework of the density-matrix renormalization group (DMRG), we study a quantum dot coupled to a superconducting nanowire with strong Rashba spin-orbit coupling. Regarding the singlet-to-doublet ``0-$\pi$'' transition that takes place when the Kondo effect is overcome by the superconducting gap, we show that the Rashba coupling modifies the critical values at which the transition occurs, favouring the doublet phase. In addition, using a generalized  Haldane's formula for the Kondo temperature $T_K$, we show that it is \emph{lowered} by the Rashba coupling. We  benchmark our DMRG results comparing them with previous numerical renormalization group (NRG) results. The excellent agreement obtained opens the possibility of studying chains or clusters of impurities coupled to superconductors by the means of DMRG.
\end{abstract}

\maketitle


\section{Introduction}\label{intro}

The interplay between magnetism and superconductivity is a fundamental topic in condensed matter physics, and plays an important role in many low-temperature phenomena, e.g. in high-temperature superconductors \cite{Dagotto94_review, Lee06_Review_HTSC, Dai2012}, inhomogeneous superconductivity\cite{Larkin64_FFLO, Fulde64_FFLO,Matsuda07_Review_FFLO}, Abrikosov vortex lattices\cite{blatter_vortex_review}, etc. Already at the microscopic level, this interplay is fascinating and complex: a single magnetic impurity [e.g., a magnetic atom or a quantum dot (QD) attached to superconducting leads] can locally break Cooper pairs and introduce single-particle localized states known as Yu-Shiba-Rusinov, or simply ``Shiba'', states inside the superconducting gap \cite{Yu65_YSR_states,Shiba68_YSR_states,Rusinov69_YSR_states}. Shiba states have been recently the focus of intense research due to their potential uses in spintronic devices and in topological superconductors hosting Majorana bound states (i.e., Majorana ``Shiba chain'' proposals)\cite{Nadj-Perdge13_Majorana_fermions_in_Shiba_chains,Klinovaja13_TSC_and_Majorana_Fermions_in_RKKY_Systems,Braunecker13_Shiba_chain}. They have been clearly seen in STM experiments in nanostructured magnetic adatom/superconductor (SC) surfaces\cite{Yazdani97_YSR_states, Ji08_YSR_states, Iavarone10_Local_effects_of_magnetic_impurities_on_SCs, Ji10_YSR_states_for_the_chemical_identification_of_adatoms, Franke11_Competition_of_Kondo_and_SC_in_molecules, Bauer13_Kondo_screening_and_pairing_on_Mn_phtalocyanines_on_Pb, Hatter15_Magnetic_anisotropy_in_Shiba_bound_states_across_a_quantum_phase_transition, Ruby15_Tunneling_into_localized_subgap_states_in_SC, Ruby_2016, Hatter2017_Scaling_of_YSR_energies, Choi2017_Mapping_the_orbital_structure_of_Shiba_states}, and in transport experiments on hybrid nanostructures made of Coulomb-blockaded quantum dots coupled to superconducting leads \cite{deFranceschi10_Hybrid_SC_QD_devices}. In these systems, the QD (physically a  carbon nanotube, a gated semiconductor, or a molecule such as C$_{60}$) acts as an artificial ``magnetic atom'' where the position of the Shiba states (also known as Andreev bound-states in this context) can be controlled with supercurrents or voltage gates. 

An important feature of Shiba physics is the existence of an experimentally accessible  spin- and parity-changing phase transition, the so called ``0-$\pi$ transition'', related to the position of the Shiba level inside the gap. For a model with a classical impurity \cite{Sakurai70}, it can be seen that the change in the occupation of a Shiba state when it crosses the Fermi level causes the collective ground state to change from a BCS-like even-parity singlet to an odd-parity doublet. When dealing with the quantum impurity case, theoretical progress both on the analytical\cite{Zittartz_1970_I,Zittartz_1970_III} and numerical side (in particular, implementations of the numerical renormalization group (NRG) method\cite{Satori92_Magnetic_impurities_in_SC_with_NRG, Sakai93_Magnetic_impurities_in_SC_with_NRG,Yoshioka98_Kondo_impurity_in_SC_with_NRG,Yoshioka00_NRG_Anderson_impurity_on_SC}), allowed to identify two competing mechanisms which operate on this transition: the Kondo effect vs the superconducting pairing potential. 
Whereas the Kondo  effect consists in the formation of a many-body singlet
in which the magnetic impurity is screened by the conduction electrons
at temperatures  $T\ll T_K$ (i.e., the Kondo temperature \cite{hewson}), the superconducting pairing potential tends to favor a Cooper-pair condensate in which the mangnetic impurity remains unscreened  due to the presence of a gap $2\Delta$ in the density of states of the superconductor around the Fermi energy $E_F$\cite{Heinrich18_Review_single_adsorbates}. 
In particular, the 0-$\pi$ transition takes places when the Shiba state (whose position inside the gap depends on the ratio $T_K/\Delta$) crosses $E_F$, something that is theoretically predicted to occur at the critical value $T_K/\Delta_\text{c} \approx 0.3$\cite{Satori92_Magnetic_impurities_in_SC_with_NRG, Sakai93_Magnetic_impurities_in_SC_with_NRG,Yoshioka98_Kondo_impurity_in_SC_with_NRG,Yoshioka00_NRG_Anderson_impurity_on_SC}.

Recently, important new questions driven by the rapidly-evolving experimental techniques have arisen in the context of Shiba physics. Among the many questions originated in the complexities of real experiments, the effect of the Rashba spin-orbit coupling (SOC) on the 0-$\pi$ transition constitutes an open problem. Here the question is: what is the effect of the Rashba SOC on, e.g., the position of the subgap states? This question was addressed recently for a classical \cite{Kim15_Classical_Shiba_states_with_Rashba_SOC} or a quantum \cite{Li18_Rashba-induced_Kondo_screening_magnetic_impurity_two-dimensional} impurity embedded in a two-dimensional (2D) superconductor. Here we are interested in the regime where the spin of the impurity is a \emph{quantum-mechanical} object in contact with a one-dimensional (1D) superconducting nanowire. As we will show,  important differences arise with respect to the 2D (classical o quantum) case. We recall that Rashba SOC is  a crucial ingredient to observe topological Majorana quasiparticle excitations in 1D systems, both in proximitized semiconductor nanowire experiments \cite{Mourik12_Signatures_of_MF,Das2012_Zero-bias_peaks_and_splitting_in_an_Al-InAs_nanowire_topological_superconductor_as_a_signature_of_Majorana_fermions, Churchill2013_Majorana_NW, Albrecht2016, Deng16_MBS_in_QD_hybrid_NW,Gül2018, Zhang2018} , and in the Majorana ``Shiba-chain'' experiments\cite{NadjPerge14_Observation_of_Majorana_fermions_in_Fe_chains,Pawlak16_Probing_Majorana_wavefunctions_in_Fe_chains,Ruby15_MBS_in_Shiba_chain,Feldman16_High_resolution_Majorana_Shiba_chain,Jeon17_Distinguishing_MZMs}.  Therefore, its effect cannot be disregarded in those experimental systems. However, due to the complicated many-body Kondo correlations that emerge already at a single quantum-impurity level, the effect of Rashba SOC on the 0-$\pi$ transition is hard to describe in detail. 

Another important question is the effect of the Rashba SOC on the Kondo temperature $T_K$. In the case of magnetic impurities in normal metals (i.e., in the absence of superconductivity), treated with the Kondo or Anderson models, the complexity of the problem has resulted  in a variety of different conclusions. Depending on the parameter regime (mainly, on the position of the localized impurity level), some authors\cite{Isaev12_Kondo_effect_in_the_presence_of_spin-orbit_coupling,Zarea12_Enhancement_of_Kondo_effect_through_Rashba_SOC} find an enhancement of $T_K$ induced by SOC, while others \cite{Malecki07_Two_Dimensional_Kondo_Model_with_Rashba_Spin-Orbit_Coupling, Zitko11_Kondo_effect_in_the_presence_of_Rashba_spin-orbit_interaction, Chen16_The_Kondo_temperature_of_a_two-dimensional_electron_gas_with_Rashba_spin–orbit_coupling,Wong16_Influence_Rashba_spin-orbit_coupling_Kondo_effect} predict a minor modification. In the case of an impurity coupled to a one dimensional metalic nanowire, it has been reported \cite{deSousa16_Kondo_effect_quantum_wire_with_spin-orbit_coupling} that the effect of SOC in the nanowire is to increase $T_K$. Finally, for an impurity in contact with a 2D superconductor with Rashba SOC \cite{Li18_Rashba-induced_Kondo_screening_magnetic_impurity_two-dimensional}, the authors report an enhancement of the screening mechanisms.   

Motivated by these questions, in this work we study a single-level QD  coupled  to a 1D superconductor with strong Rashba SOC, and study its effect on the position of the subgap states, on the $0-\pi$ transition and on the Kondo temperature. We model the QD with the Anderson impurity model with on-site interaction $U$, and assume that the superconductor is a single-channel one-dimensional (1D) nanowire subject to a local pairing term $\Delta$ and to a strong Rashba SOC. To solve this problem, we have implemented the density-matrix renormalization group (DMRG) \cite{White_DMRG_92} and have introduced a logarithmic discretization in the 1D conduction band (as is  usual in NRG implementations\cite{wilson75}). To the best of our knowledge, the DMRG method has not been applied to the Shiba-impurity problem before. Since in a 1D geometry this method is known to have a good performance with increasing number of impurities, it might offer a versatile platform to study, e.g., small 1D clusters of magnetic nanostructures coupled to superconductors. We have tested and benchmarked our results using previous works where the NRG method was used in the absence of SOC\cite{Bauer07_NRG_Anderson_model_in_BCS_superconductor, Sakai93_Magnetic_impurities_in_SC_with_NRG}, with excellent agreement, showing that the DMRG reaches essentially the same degree of accuracy. These encouraging results pave the way to implement DMRG as a reliable alternative to describe many-body physics of subgap states induced by magnetic impurities. 

Our results show that the Rashba SOC in a 1D setup does not qualitatively affect the phase diagram of the 0-$\pi$ transition, affecting it only at a quantitative level through a modification of the effective local (i.e., at the site of the QD) density of states at the Fermi level $\rho_0$. This also leads us to conclude that in a 1D geometry the Rashba SOC has \emph{detrimental} effects on $T_K$. 

This paper is organized as follows. We begin in Sec. \ref{model} by describing the 1D model of a QD coupled to a superconductor nanowire with Rashba SOC. In Sec. \ref{Logarithmic} we focus on the DMRG and, in particular, on the implementation of the  logarithmic-discretization procedure used to map our model onto an effective ``Wilson chain'' Hamiltonian\cite{wilson75}. In Sec. \ref{res} we present our DMRG results, focusing mainly on the position of the subgap states and the 0-$\pi$ transition as a function of the Rashba SOC parameter. Our results are computed both at and away the particle-hole symmetric point of the Anderson model.  Lastly, we devote Sec. \ref{conclusions} to present a summary and discuss future perspectives.

\section{Theoretical Model}\label{model}

We describe a single-level QD hybridized with a superconducting lead by means of the Anderson impurity model $H = H_\text{SC} + H_{d}$. Here, the term $H_\text{SC}$ describes a single-channel BCS superconductor
with a Rashba SOC term: 
\begin{align}
H_\text{SC}&= \sum_{k} 
\left[\sum_\sigma \epsilon_0\left(k\right) c^{\dagger}_{k \sigma} c_{k \sigma}+\Delta \left(c^{\dagger}_{k \uparrow} c^{\dagger}_{-k \downarrow} + \text{H.c.}\right)\right.
\nonumber\\ 
&\left.-\sum_{\sigma,\sigma^\prime} \hbar \alpha_R  k \left(c^{\dagger}_{k \sigma} \hat{\sigma}^y_{\sigma,\sigma^\prime} c_{k \sigma^\prime} \right)\right],
\label{H_SC}
\end{align}
where  $c_{k\sigma}$ is the annihilation operator of a fermionic quasiparticle with momentum $k$ and spin projection $\sigma$ along the $\hat{z}$-axis, and is defined so that the usual commutation relation $\{c_{k\sigma},c^\dagger_{k^\prime 
\sigma^\prime}\}=\delta_{k,k^\prime}\delta_{\sigma,\sigma^\prime}$ holds.
The quantity $\epsilon_0\left(k\right)$ is the dispersion relation of the quasiparticles in  the 1D conduction band in the absence of both Rashba SOC and pairing interaction, and is taken with respect to $E_F$.
The BCS pairing potential $\Delta$ induces a superconducting ground state, and opens a SC gap of size 2$\Delta$ around $E_F$ in the spectrum of quasiparticles\cite{Tinkham_Introduction_to_superconductivity}. 
For simplicity, here we have not considered the usual self-consistent equation for the BCS order parameter, and therefore our results are limited to the regime $T\ll T_\text{c}$, with $T_\text{c}$ the superconducting critical temperature. Finally, $\alpha_R$ is the Rashba SOC parameter (note that $\alpha_R$ has units of velocity), generated by the breaking of the inversion symmetry, and $\hat{\sigma}^y$ is the Pauli matrix acting on spinor space. 

This model could represent the situation in, e.g., a semiconducting quantum wire proximitized by a nearby bulk superconductor  (Pb or Al), as used recently in Majorana experiments\cite{Mourik12_Signatures_of_MF,Das2012_Zero-bias_peaks_and_splitting_in_an_Al-InAs_nanowire_topological_superconductor_as_a_signature_of_Majorana_fermions, Churchill2013_Majorana_NW,Albrecht2016,Deng16_MBS_in_QD_hybrid_NW,Gül2018, Zhang2018}
We stress that the assumption of a 1D superconductor is not essential for the implementation of the numerical techniques presented in this work. Higher-dimensional geometries, such as magnetic impurities in 2D SCs, are also possible to describe but we defer these studies for future works.

The term $H_d$ describes a single-level QD with strong local Coulomb repulsion $U$, coupled to the SC:
\begin{align}
H_{d} &= \sum_{\sigma} \epsilon_d n_{d \sigma}+U n_{d \uparrow}n_{d \downarrow}+\sum_{k\sigma} \frac{V}{\sqrt{N}} \left(d^{\dagger}_{\sigma} 
c_{k \sigma} + \text{H.c.}\right), \label{H_imp}
\end{align}
where $d^{\dagger}_{ \sigma}$ creates an electron with spin proyection $\sigma$ in the QD, and $n_{d\sigma}=d^\dagger_\sigma 
d_\sigma$ is the number of fermions. The parameter $\epsilon_d$ is the energy level of the dot, which is assumed to be tuned by means of external gate voltages, and $U$ is the local Coulomb repulsion. The parameter $V$ is the hybridization hopping amplitude between the QD and the SC nanowire. For later use, it is convenient to define here the \emph{effective} hybridization parameter $\Gamma_0=V^2\rho_0 \pi$, where $\rho_0$ is the density of states at the Fermi level. 

To gain insight into the physical aspects of this Hamiltonian, we can assume the system in a particle-hole symmetric situation (i.e., $\epsilon_d=-U/2$) and $\Gamma_0 \ll U$. Under such conditions, the QD acts as an effective spin-1/2 impurity,  with a frozen occupation number in the subspace $n_{d \sigma} \simeq 1/2$. In the absence of SC pairing (i.e., $\Delta=0$), the conduction-band electrons near $E_F$ tend to screen this effective spin-1/2 impurity and create Kondo correlations which eventually give rise to the many-body ``Kondo singlet'' characterized by an energy scale $T_K\sim D \exp{\left[-\pi U/8 \Gamma_0\right]}$\cite{Haldane_1978}. However, in the presence of the SC gap, due to the lack of quasiparticles in the energy region $2\Delta$ around $E_F$, the screening mechanism fails if $T_K \ll \Delta$ and the QD remains unscreened. This is the essence of the ``0-$\pi$'' transition. 

Before implementing the numerical solution of this model, it is convenient first to introduce a unitary transformation in spinor space in order to eliminate the Rashba SOC from $H_\text{SC}$, i.e., $\left(\tilde{c}_{k +}, \tilde{c}_{k -}\right)^T= \hat{U} \left(c_{k \uparrow}, c_{k \downarrow}\right)^T$, where the unitary transformation $\hat{U}$ is a $\frac{\pi}{2}$-rotation in spinor space  around the $\hat{x}$ axis: $\hat{U}=e^{i \frac{\pi}{4} \hat{\sigma}_x}/\sqrt{2}$. In this new basis, the transformed Hamiltonian $\tilde{H}_\text{SC}=\hat{U}^\dagger H_\text{SC} \hat{U}$ explicitly writes  
\begin{align}
\tilde{H}_\text{SC}&= \sum_{k} \left[\sum_{h=\pm} \epsilon_{h}\left(k\right) \tilde{c}^{\dagger}_{k h} \tilde{c}_{k h}+ \Delta \left(\tilde{c}^{\dagger}_{k +} \tilde{c}^{\dagger}_{(-k) -} +  \text{H.c.} \right)\right], \label{H_rot}
\end{align} 
where we have defined the new band dispersion $\epsilon_{h}\left(k\right) \equiv \epsilon_0\left(k\right) + h \alpha_Rk \hbar$, with $h=\pm$ playing the role of an effective ``up'' or ``down'' spin projection along the $\hat{y}$-axis. The same transformation can be implemented for the QD term, $\tilde{H}_d=\hat{U}^\dagger H_d \hat{U}$. Explicitly:
\begin{align}
\tilde{H}_{d} &= \sum_{h=\pm}\epsilon_d \tilde{n}_{d h}+U \tilde{n}_{d +}\tilde{n}_{d -}+\sum_{k,h=\pm} \frac{V}{\sqrt{N}} \left(\tilde{d}^{\dagger}_{h} \tilde{c}_{k h} + \text{H.c.}\right), \label{H_rotimp}
\end{align}
where the new impurity operators are $\left(\tilde{d}_{+}, \tilde{d}_{-}\right)^T= \hat{U} \left(d_{\uparrow}, 
d_{\downarrow}\right)^T$. 

Note that in the transformed Hamiltonian $\tilde{H}=\tilde{H}_\text{SC}+\tilde{H}_d$, the Rashba SOC term has been eliminated and is now completely encoded in the new dispersion relation $\epsilon_{kh}$. Moreover, this transformed Hamiltonian is a one-channel Anderson Hamiltonian. This is not a peculiarity of 1D: as shown in Ref. \onlinecite{Zitko11_Kondo_effect_in_the_presence_of_Rashba_spin-orbit_interaction}, the single-orbital Anderson impurity model is always effectively a single-channel problem, independently of dimensionality and of the type of conduction band.  

Following Malecki \cite{Malecki07_Two_Dimensional_Kondo_Model_with_Rashba_Spin-Orbit_Coupling}, we assume a quadratic dispersion $\epsilon_0\left(k\right)=\hbar^2k^2/2m^*-\mu$, with $m^*$ the renormalized mass of the band quasiparticles and $\mu$ the chemical potential. With this, the Fermi energy is $E_F=\mu$. We obtain a modified Fermi wavevector and Fermi velocity due to the Rashba SOC:
\begin{align}
k_{Fh}&=k^0_F\sqrt{1+\frac{\epsilon_R}{\mu}}-hk_R\hbar,\\
v_{Fh}&=\frac{1}{\hbar} \left. \frac{\partial \epsilon_{kh}}{\partial k}\right|_{k=k_{Fh}}=v^0_F\sqrt{1+\frac{\epsilon_R}{\mu}} \label{eq:vFh},
\end{align}
where $k^0_F=\sqrt{2m^*\mu}/\hbar$ and $v^0_F=\hbar k^0_F/m^*$ are, respectively, the Fermi wavevector and the Fermi velocity in the absence of Rashba SOC, and where we have defined a ``Rashba momentum'' $k_R=m^*\alpha_R/\hbar$, and a Rashba energy $\epsilon_R=m^*\alpha_R^2/2$, so that $\epsilon_R/\mu=\left(\alpha_R/{v^0_F}\right)^2$. In the following, when refering to the effects of the Rasbha SOC, we will alternatively refer to the Rashba energy $\epsilon_R$ or to the Rashba coupling $\alpha_R$. 

Since in a 1D geometry the density of normal states at the Fermi energy is obtained from the expression $\rho_0\left(\epsilon_R\right) = 1/\left(2 \pi v_{Fh}\right)$, from Eq. (\ref{eq:vFh}) we can obtain the expression of the density of states \textit{modified} by the effect of the Rashba SOC 
\begin{align}
\rho_0\left(\epsilon_R\right/\mu)&=\frac{\rho_0}{\sqrt{1+\frac{\epsilon_R}{\mu}}}
\label{eq:rho0R}.
\end{align}
Therefore, in this 1D case the effect of the Rashba SOC appears  \textit{only} through a modification of the density of states of the conduction band, for the purposes of this work\cite{note1}. 

In what follows, we will assume that the Fermi level is far from the bottom of the band, which we assume located at energy $\epsilon=-D$, and therefore we linearize the 1D spectrum in a window of energy $2D$ around $E_F$, i.e., $\epsilon_h\left(k\right)\simeq v_{Fh}k$. This amounts to replacing the original band by a symmetric, half-filled flat band with a constant density of states in the region $E_F-D < \epsilon < E_F+ D$. This is the most important approximation in our work, which nevertheless is the standard case in most NRG studies (as we will see in  Sec. \ref{Logarithmic}, it considerably simplifies the implementation of the logarithmic discretization method). In addition, this approximation imposes the condition $\mu=D$ in  Eq. \ref{eq:rho0R} (i.e., half-filled band), and therefore the decrease in the density of states at $E_F$ can be interpreted in terms of a modified (broader) effective conduction band due to the effect of the Rashba SOC, i.e., $D\rightarrow D\left(\epsilon_R\right)\equiv D\sqrt{1+\frac{\epsilon_R}{D}}$. In this way, the product $2D\left(\epsilon_R\right)\rho_0\left(\epsilon_R\right)$ remains constant and the number of electrons in the effective conduction band is preserved. We stress that this  is  a \emph{generic} property of a Rashba-coupled 1D band, and is independent of the above approximation. Consequently, it is easy to see that the effective hybridization is renormalized to lower values:
\begin{equation}
 \Gamma\left(\epsilon_R\right/D)=\frac{\Gamma_0}{\sqrt{1+\frac{\epsilon_{R}}{D}}}.
 \label{newGamma}
\end{equation}
This result is consistent with Refs. \onlinecite{Zitko11_Kondo_effect_in_the_presence_of_Rashba_spin-orbit_interaction, Chen16_The_Kondo_temperature_of_a_two-dimensional_electron_gas_with_Rashba_spin–orbit_coupling},
where the same conclusion was obtained using, respectively, the NRG and Monte Carlo approaches.

\section{Logarithmic discretization and DMRG}

\label{Logarithmic}

Having established that the effect of the Rashba SOC enters essentially through a renormalized density of states, we now focus on the implementation of the DMRG method in order to obtain the ground-state properties of the system. An important feature of this problem is that the BCS Hamiltonian (\ref{H_rot}) does not preserve the number of particles: the presence of the pairing term $\sim \Delta \left(\tilde{c}^{\dagger}_{k +} \tilde{c}^{\dagger}_{(-k) -} +  \text{H.c.} \right)$ changes the number of particles of a $N$-particle state  by $\pm 2$. Note, however, that the fermion parity, i.e. $P=(-1)^N$, is a conserved quantity which can be used to classify the different many-body states in the Hilbert space when implementing DMRG.

One important aspect of Shiba systems is the exponential localization of subgap states characterized by a localization length $\xi$. A rough estimation of $\xi$ in our case can be obtained assuming, for  simplicity, a classical (instead of a quantum) spin. Following Ref. \onlinecite{Balatsky_2006},  for the simplest case of isotropic scattering the Shiba state is localized around the impurity with localization length
$\xi=\xi_{0}/\left|\sin\left(2\delta_{0}\right)\right|$,
where $\xi_{0}$ is the coherence length of the BCS superconductor,  defined as $\xi_{0}\simeq\hbar v_{F}/\Delta$, and $\delta_{0}$ is the $s$-wave phase shift due to the magnetic scattering with the impurity. From Eq.  (6.10) in Ref. \onlinecite{Balatsky_2006}, the relation between $E_b$, the energy of the Shiba state within the gap,  and the phase shift is 
$\frac{E_b}{\Delta} =\cos\left(2\delta_{0}\right)$, and therefore, we can write the localization length as
\begin{align}
\xi & =\frac{\xi_{0}}{\sqrt{1-\frac{E_b^2}{\Delta^2}}}.
\end{align}
From this expression, it is easy to realize that the localization length
of the Shiba level diverges as its energy gets close to the superconductor gap edge (i.e., $E_b/\Delta \rightarrow 1$). This is particularly problematic for real-space methods such as DMRG, which can reach system sizes of up to $L_\text{max} \sim 300$ sites, depending on the implementation. This means that eventually, the localization length will be $\xi \gg L_\text{max} $, and considerable errors arising from finite-size effects will appear. 

As mentioned before, the case of a single impurity coupled to a SC host \emph{without} Rashba SOC has been studied in previous works by means of the 
NRG method
\cite{Satori92_Magnetic_impurities_in_SC_with_NRG, Sakai93_Magnetic_impurities_in_SC_with_NRG,Yoshioka98_Kondo_impurity_in_SC_with_NRG,Yoshioka00_NRG_Anderson_impurity_on_SC,  Bauer07_NRG_Anderson_model_in_BCS_superconductor,Zitko11_Kondo_effect_in_the_presence_of_Rashba_spin-orbit_interaction, Zitko16}. One crucial step in NRG implementations corresponds to the logarithmic discretization procedure of the conduction band, and the subsequent mapping of the Hamiltonian onto a Wilson chain Hamiltonian. 
Here, although in principle the DMRG \emph{does not} require such mapping, we will adopt it in order to deal with the extremely large subgap-state localization length, which generically exceeds the maximal system sizes allowed by our computational resources. Therefore, following the abovementioned references, we implement a logarithmic discretization, which effectively maps the original Hamiltonian $\tilde{H}_\text{SC}$ defined in $k$-space, onto an effective  one-dimensional semi-infinite chain (i.e., the Wilson chain). Since we follow a standard technique, we do not provide here the details of this derivation and refer the reader to Refs. \onlinecite{Bulla08_NRG_review,wilson75, Satori92_Magnetic_impurities_in_SC_with_NRG, Sakai93_Magnetic_impurities_in_SC_with_NRG,Yoshioka98_Kondo_impurity_in_SC_with_NRG, Yoshioka00_NRG_Anderson_impurity_on_SC}, where the method is well explained. Applying the logarithmic discretization the effective Wilson chain Hamiltonian is obtained:

\begin{widetext}
\begin{align}
\bar{\tilde{H}} &=  \bar{U} \tilde{n}_{d +} \tilde{n}_{d -} + \sum_{h} \left[\bar{\epsilon}_d \tilde{n}_{d h} +  \bar{V} 
\left(\tilde{d}^{\dagger}_{h} \tilde{f}_{1 h} + \text{H.c.}\right) \right]+ \sum_{n=0}^\infty \left[ \sum_{h=\pm} 
\left(\bar{\gamma}_{n} \tilde{f}^{\dagger}_{n h} \tilde{f}_{n+1 h} + \text{H.c.}\right)+\bar{\Delta} 
\left(\tilde{f}^{\dagger}_{n +} \tilde{f}^{\dagger}_{n -} + \text{H.c.} \right) \right], \label{H_DMRG}
\end{align}
\end{widetext}
where the bar indicates dimensionless quantities expressed in units of $D$ (e.g., $\bar{\Delta}\equiv \Delta/D$). The  
index $n$ (interpreted here as the effective ``site'' in the Wilson chain) corresponds to the $n$-th energy shell 
$\varLambda^{-\left(n+1\right)}<|\bar{\epsilon}|<\varLambda^{-n}$ in the logarithmically-discretized conduction band, with 
$\varLambda>1$ the discretization parameter. Consistently, the operator $f_{nh}^\dagger$ creates a fermion in that energy 
shell. The effective  ``hopping'' parameter $\bar{\gamma}_{n}$ acquires the form\cite{wilson75,Bulla08_NRG_review}:

\begin{equation}
 \bar{\gamma}_{n } = \sqrt{1+\frac{\epsilon_R}{D}} \frac{(1+\varLambda^{-1})(1-\varLambda^{-n-1})}{2 
\sqrt{(1-\varLambda^{-2n-1})} \sqrt{(1-\varLambda^{-2n-3})}} \varLambda^{-n/2},
\end{equation}
which is standard, except for the extra renormalization factor $\sqrt{1+\frac{\epsilon_R}{D}}$ due to the effect of the Rashba SOC. This renormalization of the effective hopping parameter is directly related to the change in the conduction band width $D\left(\epsilon_R\right)\equiv D\sqrt{1+\frac{\epsilon_R}{D}}$ described in the previous Section.   

The Hamiltonian (\ref{H_DMRG}) has been solved using the DMRG method\cite{White_DMRG_92}, for various values of the parameter $U/\Delta$. We have calculated the ground state (GS) energy and also the spectral function $\rho_d\left(\omega\right)$ of the QD. To the best of our knowledge, the DMRG has not been used before to solve this kind of system, which has been treated using mainly the NRG method \cite{Satori92_Magnetic_impurities_in_SC_with_NRG, Sakai93_Magnetic_impurities_in_SC_with_NRG,Yoshioka98_Kondo_impurity_in_SC_with_NRG,Yoshioka00_NRG_Anderson_impurity_on_SC,  Bauer07_NRG_Anderson_model_in_BCS_superconductor,Zitko11_Kondo_effect_in_the_presence_of_Rashba_spin-orbit_interaction, Zitko16}, perturbation theory in $U$ \cite{Zonda16_Perturbation_theory_of_SC_0_Pi_transition}, or equations of motions \cite{Li18_Rashba-induced_Kondo_screening_magnetic_impurity_two-dimensional}.

In all cases, we have used a discretization parameter $\varLambda=2$, larger values are not convenient since they would tend to concentrate a large number of discrete frecuencies near $E_F$, where the SC gap suppresses the density of states \cite{Bauer07_NRG_Anderson_model_in_BCS_superconductor}. This allows us to work with realistic values of $\Delta$ (we consider a typical value for the bandwidth $D\simeq $ 1 eV, and for the SC gap $\Delta\simeq 1$ meV, so typically $\bar{\Delta}\simeq 10^{-3}$), and also with much smaller values of $\Delta$ when testing the universality of the model (see Section \ref{Universality} below). The presence of the SC gap is actually beneficial for the  DMRG method \cite{DMRG_review}. A technical point here is that as a consequence of the discretization, the density of conduction states at the Fermi level $\rho_{0,\varLambda}\left(0\right)$ decreases with respect to the continuum limit $\varLambda \rightarrow 1$. We have calculated $\rho_{0,\varLambda}\left(0\right)$ numerically for the Wilson chain without the impurity and in the absence of Rashba SOC, and determined the hybridization $V_\varLambda$ of the QD with the first site of the chain from the condition $\Gamma_0=\pi \rho_{0,\varLambda} V_\varLambda^2$.

\section{Results}
\label{res}
 
When the QD is connected to the  superconductor, the multiple Andreev reflections of fermions at the QD-SC nanowire interface  give rise to localized Shiba or Andreev bound states. As already mentioned, the GS of the system can be either a singlet, in which case the fermion parity is even, or a doublet, in which case the parity is odd (see Ref. \onlinecite{Zitko16}). The energy difference between the odd-parity and even-parity GSs  gives the energy of the subgap Shiba level $E_{b}$:
\begin{equation}
 E_{b} = \pm\left(E_\text{o} - E_\text{e}\right),
 \label{eq:Shiba}
\end{equation}
where $ E_\text{e(o)} $ is the GS energy for the system in the subspace with even (odd) number of particles.  The $\pm$ signs appear due to the intrinsic particle-hole symmetry of the BCS Hamiltonian: each quasi-particle eigenstate $\psi\left(\epsilon \right)$ with energy $+\epsilon$ is related by charge-conjugation to  a ``partner'' quasi-hole state $\psi^\dagger\left(-\epsilon \right)$ at energy $-\epsilon$. Subgap states are not the exception and therefore, finite-energy Shiba states \emph{must} appear in pairs symmetrically located around the $E_F$.

\subsection{Symmetrical point}
\begin{figure}[t]
\includegraphics[width=8.5cm]{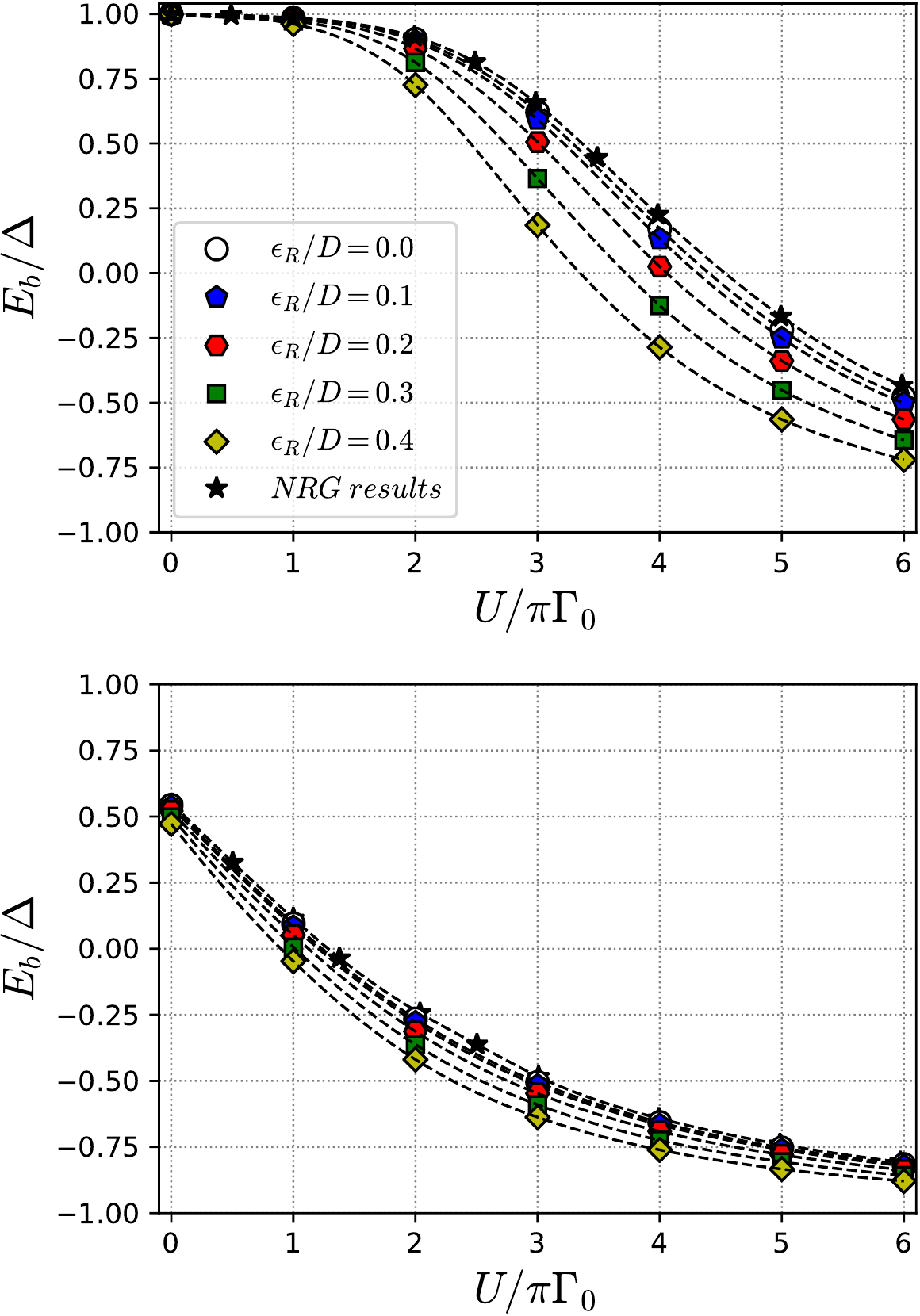}
\caption{(Color online) Energy (in units of the SC gap $\Delta$) of one branch of the Shiba states as a function of $U/\pi 
\Gamma_0$ for $\Delta=0.001$ (top) and  $\Delta=0.06$ (bottom). Starting from $U/\pi \Gamma_0=0$, when the energy crosses the line $E_b / \Delta =0$, the transition from a Kondo singlet to a doublet occurs. We remark that there is another branch symmetrically located respecto to zero energy, not shown for clarity. The NRG results for zero Rashba SOC coupling are from Ref. 
\onlinecite{Bauer07_NRG_Anderson_model_in_BCS_superconductor}}
\label{Energy_Shiba}
\end{figure}

We concentrate first on the electron-hole symmetric point of the Anderson model, i.e.,  $\epsilon_d=-U/2$. In Fig. 
\ref{Energy_Shiba} we show the energy of the Shiba states as a function of $U/\pi \Gamma_0$ for different values of 
$\epsilon_R$ and for different values of $\Delta$. For clarity, here we plot only the ``+'' Shiba-state branch in Eq. 
(\ref{eq:Shiba}). The singlet-to-doublet  transition takes place when $E_b$ crosses zero energy. We can observe that the effect of the Rashba SOC is to shift the transition point to lower $U/\pi \Gamma_0$, thus
favoring the transition to a doublet induced by the superconducting pairing interaction. Note that the effect of 
the Rashba SOC is more important for smaller SC gap $\Delta$. As a benchmark for validating our method, we have checked 
that the curves for $\epsilon_R=0$ match those reported previously calculated with NRG 
\cite{Bauer07_NRG_Anderson_model_in_BCS_superconductor, Zitko16, Yoshioka00_NRG_Anderson_impurity_on_SC}. As shown in Fig. 
\ref{Energy_Shiba}, for all values of  $U/\pi \Gamma_0$ and/or the SC gap $\Delta$, the agreement is excellent. 

\begin{figure}[h]
\includegraphics[width=8.5cm]{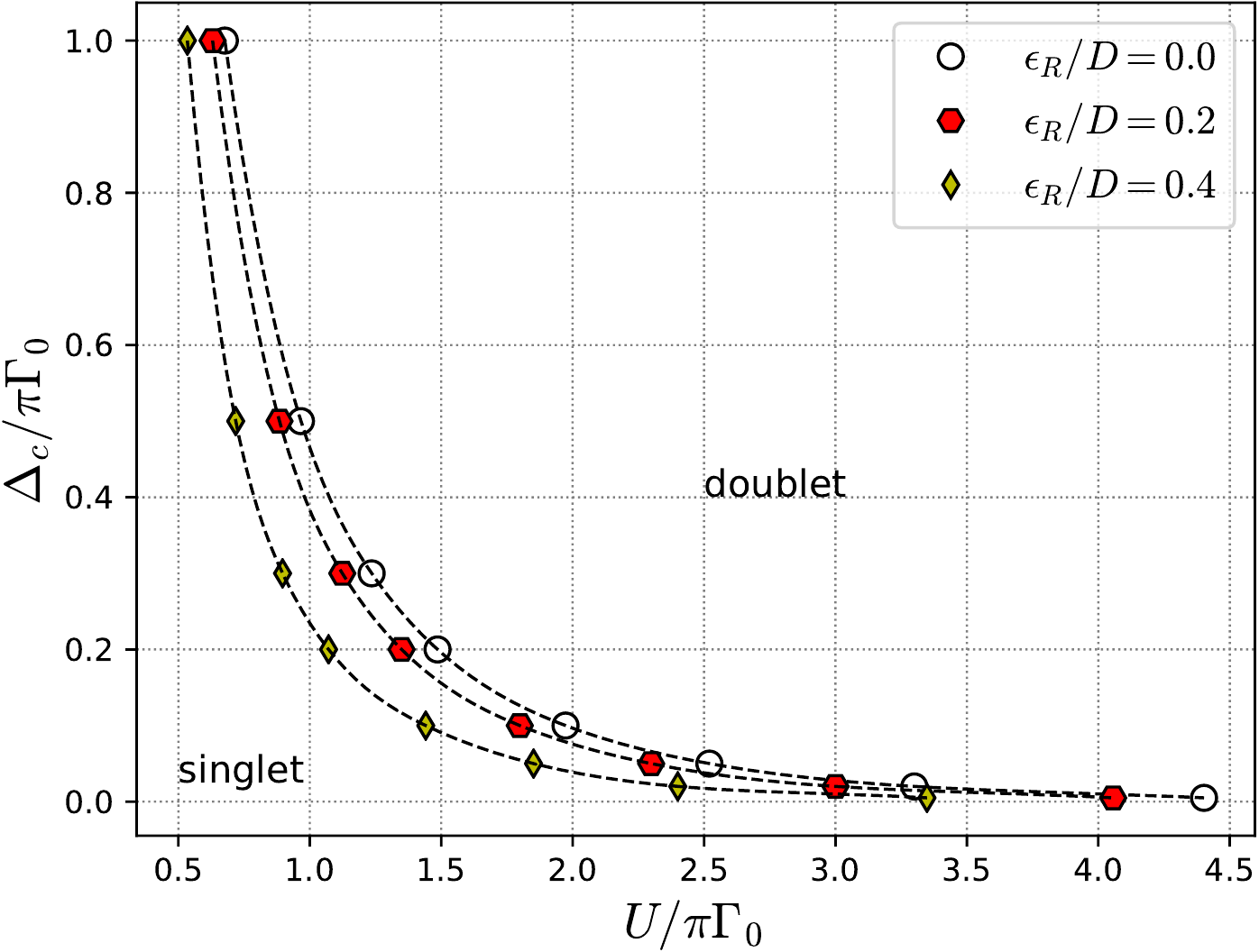}
\caption{(Color online) Phase diagram of the model showing the critcal gap as a function of $U/\pi \Gamma_0$ for different values of the Rashba energy.}
\label{diagrama_fases}
\end{figure}

In Fig. \ref{diagrama_fases} we show the critical gap $\Delta_\text{c}$ as a function of $U$, both scaled by $\pi 
\Gamma_0$. This type of diagram is similar to that shown by Bauer {\it{et al.}} (see Fig. 9 in Ref. 
\onlinecite{Bauer07_NRG_Anderson_model_in_BCS_superconductor}), and in addition we show the effect of the Rashba SOC. The 
curves $\Delta_\text{c}/\pi \Gamma_0$ vs $U/\pi \Gamma_0$ indicate the boundary between the singlet and doublet regions in 
parameter space. We see that the region with a doublet GS expands as the Rashba SOC increases (i.e., the curves are lowered 
and shifted to the left as $\alpha_R$ increases). Note that  for smaller values of $U/ \pi\Gamma_0$ (i.e., smaller than 
$\sim$ 0.63 for $\alpha_R =0$), the transition dissappears, or equivalently $\Delta_\text{c}\rightarrow \infty$. This can 
be explained by the fact that for $U/ \pi\Gamma_0\ll 1$, quantum fluctuations of the charge inhibit the formation of a net 
magnetic moment in the QD and therefore the system never reaches the doublet GS phase. The point at which 
$\Delta_\text{c}\rightarrow \infty$ also shifts to lower values for finite values of $\alpha_R$.       

In Ref. \onlinecite{Bauer07_NRG_Anderson_model_in_BCS_superconductor}, the $0-\pi$ transition was addressed varying the 
parameter $U/\pi \Gamma_0$ and/or the SC gap $\Delta$. Assuming a fixed value of $\Delta$,  the parameter $U/\pi \Gamma_0$ 
is indeed a good parameter that allows to explore the quantum phase diagram. Intuitively,  a small value of $U/\pi 
\Gamma_0$ implies a large amount of charge fluctuations (and therefore, a non-magnetic regime) in the QD. Then, the QD is 
not able to ``break'' the Cooper pairs and  consequently the  GS of the system is a singlet adiabatically connected to the 
BCS state. For moderate values of $U/\pi \Gamma_0$ such that $T_K\sim D e^{-\pi U/8\Gamma_0}\gg \Delta$, the BCS state 
smoothly evolves into a many-body Kondo singlet, and throughout this evolution, the GS remains in the singlet subspace. 
Finally, for larger values of $U/\pi \Gamma_0$ such that $T_K\ll \Delta$, the QD develops a well-defined $S=1/2$ magnetic 
moment which cannot be screened due to the quasiparticle gap, and the GS becomes a doublet. 

The effect of the Rasbha SOC term ontop of the above physical picture occurs through a modification of $\rho_0\rightarrow 
\rho_0\left(\epsilon_R\right)$ [see Eq. (\ref{eq:rho0R})], which is actually always lowered as the Rashba SOC increases, in 
our 1D case. As we will analize in detail in Sec. \ref{Universality}, this results in a weakening of the Kondo effect, 
consequently the critical $\left(U/\pi \Gamma_0\right)_\text{c}$ is shifted to lower values. Indeed, as we have seen from the analysis of Figs. \ref{Energy_Shiba} and \ref{diagrama_fases}, the effect of the Rashba SOC on the host has the effect of favouring the doublet phase. Then, if the Rashba SOC coupling could be tuned as 
a parameter in the Hamiltonian, as it happens in semiconductors and interfaces coupled to gate potentials 
\cite{Guido_18_Tuning_Rashba_spin_orbit_coupling_in_homogeneous_semiconductor_nanowires, 
Iorio_2019_Vectorial_Control_of_the_Spin_Orbit_Interaction_in_Suspended_InAs_Nanowires,
Herranz_15_Engineering_two_dimensional_superconductivity_and_Rashba_SOC, 
Direct_Rashba_spin-orbit_interaction_in_Si_and_Ge_nanowires_with_different_growth_directions}, this mechanism could be used 
for the \textit{in-situ} control the 0-$\pi$ transition.

When refering to experiments, it is worth noting that $D$ is not a relevant experimental parameter, and a more important quantity is $\mu$ which  essentially determines the filling of the band(s), 
and in semiconductor nanowires can be tuned with voltage gates
\cite{Das2012_Zero-bias_peaks_and_splitting_in_an_Al-InAs_nanowire_topological_superconductor_as_a_signature_of_Majorana_fermions}. Since we have imposed $\mu=D$ in order to have a half filled band, we can take as a reference the values of the chemical potential in experiments. In Ref.  \onlinecite{Das2012_Zero-bias_peaks_and_splitting_in_an_Al-InAs_nanowire_topological_superconductor_as_a_signature_of_Majorana_fermions} experiments in InAS nanowires report values of $\mu$ up to  $\sim 0.2$  meV, and spin-orbit energies of $\epsilon_R=75$  $\mu\text{eV}$, so in that case $\epsilon_R/\mu \sim 0.375$, in accordance with the values we have used for our calculations, but it is important to keep in mind that those experimental values might vary with the chemical potential. On the other hand, much larger Rashba spin-orbit energies $\epsilon_R\simeq 6.5$ meV have been found experimentally in InAs nanowires \cite{Wang_2017, Kammhuber2017}.

\subsection{Spectral function} 

Using the correction vector scheme of DMRG presented by Nocera and Alvarez\cite{Nocera_PhysRevE.94.053308}, we have  
calculated the spectral function at  the QD as
\begin{align}
 \rho_d\left(\omega\right) = & - \frac{1}{\pi} \text{Im} \left\{ \langle \psi_\text{GS} | d \left(\frac{1}{\omega + i \eta 
+ H - E_\text{GS} } \right) d^\dagger |\psi_\text{GS}  \rangle +\right. \nonumber \\
   & \left. \langle\psi_\text{GS} | d^\dagger \left( \frac{1}{\omega + i \eta - H - E_\text{GS} } \right) d |\psi_\text{GS} 
 \rangle \right\}
\label{eq:spectral}  
\end{align}
where here $\eta$ is a small broadening parameter that is introduced to avoid the poles of the Green's function in the real 
axis, and the GS energy $E_\text{GS}$ can be either $E_\text{e}$ or $E_\text{o}$ depending on the specific values of 
parameters $\Delta,U/\pi\Gamma_0$ and $\alpha_R$. We use a frequency-dependent 
$\eta=\eta(\omega)$ to increase the accuracy of the plot:
\begin{align}
\eta\left(\omega\right)&=\begin{cases}
\Delta/20 &\text{if }\left|\omega\right|<\Delta,\\ 
Ae^{\omega}+B &\text{if }\Delta<\left|\omega\right|< 0.2D,\\
0.15D &\text{if } 0.2D<\left|\omega\right|<D,
\end{cases}
\label{eq:eta}
\end{align}
The constants $A$ and $B$ are chosen so that $\eta(\omega)$ is a continuous function.

\begin{figure}[htb]
\includegraphics[width=8.5cm]{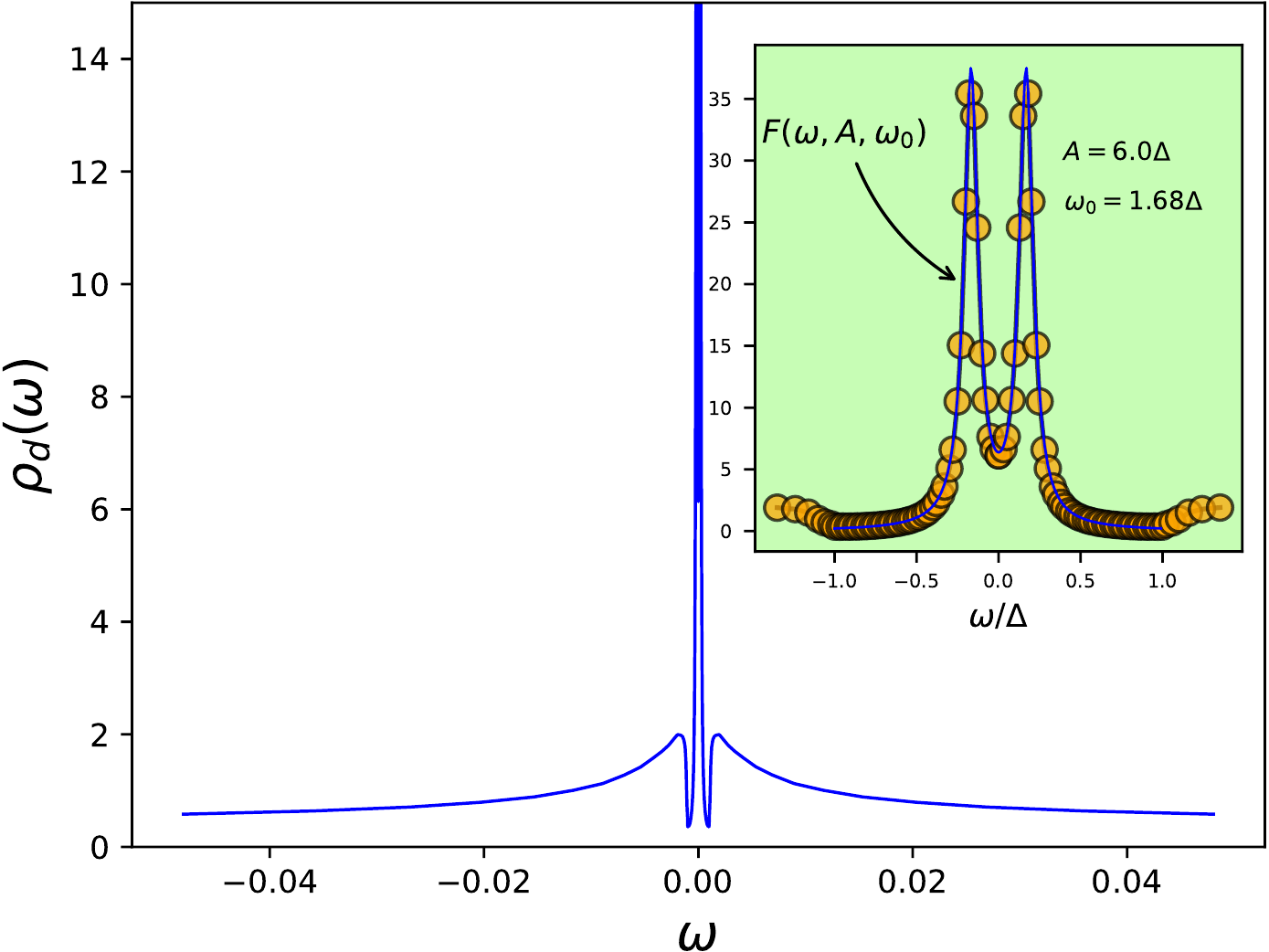}
\caption{(Color online) Spectral function $\rho_d\left(\omega\right)$ of the impurity for $U/(\pi \Gamma_0)=4, \Delta=0.001$ and $\epsilon_R=0$. Inset: zoom inside the gap to visualize the two Shiba peaks fitted by $F(\omega,\omega_0,A)=L(w,w_0,A)+L(-w,-w_0,A)$. The energy of the Shiba peaks inside the gap perfectly matches that plotted in Fig. \ref{Energy_Shiba}}
\label{spectral}
\end{figure}

In Fig. \ref{spectral} we show the calculated $\rho_d\left(\omega\right)$  for the parameters indicated in the caption, for which the system is in the singlet phase. All the expected features can be observed with clarity: the SC energy-gap of width $2\Delta$  and the Shiba states, which clearly appear as two peaks inside the gap. Each Shiba peak 
inside the gap can be described with a Lorentzian function $L(w,A,w_0)=\frac{\eta A}{\pi(\eta^2+(\omega-\omega_0)^2)}$, where  $\omega_0$ and $A$ are fitting parameters controlling, respectively, the center of the peak and its spectral weight. The parameter $\eta$ is the width of the Lorentzian and is the same function defined in Eq. (\ref{eq:eta}) and used in the correction vector calculations for each $\omega$ of the spectral function in Eq. (\ref{eq:spectral}).  As an internal sanity check, we have verified that the center of the Shiba peaks $\omega_0$ match (within the DMRG numerical precision) the corresponding values of $E_b$ obtained from Eq. (\ref{eq:Shiba}). 

With respect to the spectral weights obtained within this scheme, we do not show the results here but we mention that for $\alpha_R = 0$ our calculations agree very well with those reported in previous works, e.g., specifically with the weights appearing in Fig. 3 in Ref.\onlinecite{Bauer07_NRG_Anderson_model_in_BCS_superconductor}. The most important feature of the spectral weight is that at the 0-$\pi$ transition it has an abrupt discontinuity and its value, coming from the singlet phase, is reduced to a half on the doublet phase. This discontinuity is due to the abrupt change in the degeneracy of the GS from $g=1$ (singlet) to $g=2$ (doublet), as explained in Refs. 
\onlinecite{Sakai93_Magnetic_impurities_in_SC_with_NRG,Yoshioka98_Kondo_impurity_in_SC_with_NRG}, and consequently, it is a universal feature which is preserved at finite $\alpha_R$.

\subsection{Away from the symmetrical point}
\label{assymetry}

It has already been  established \cite{Bauer07_NRG_Anderson_model_in_BCS_superconductor} that deviations from the electron-hole symmetrical point $\epsilon_d=-U/2$ (for fixed $\Delta$, $U$ and $\Gamma_0$) can also trigger  the 0-$\pi$ transition. This situation is experimentally more relevant, since in many cases the energy level of the impurity $\epsilon_d$ cannot be controlled, and therefore it would be very rare to find the system ``self-tuned'' at the symmetric point. Moreover, in QDs or nanowires where $\epsilon_d$ can in principle be controlled by the gate potential, the asymmetry can be tuned \textit{in situ}, providing an additional ``knob'' to access the  0-$\pi$ transition. Therefore, describing the effects of electron-hole asymmetry is intrinsecally and experimentally relevant.

\begin{figure}[h]
\includegraphics[width=8.5cm]{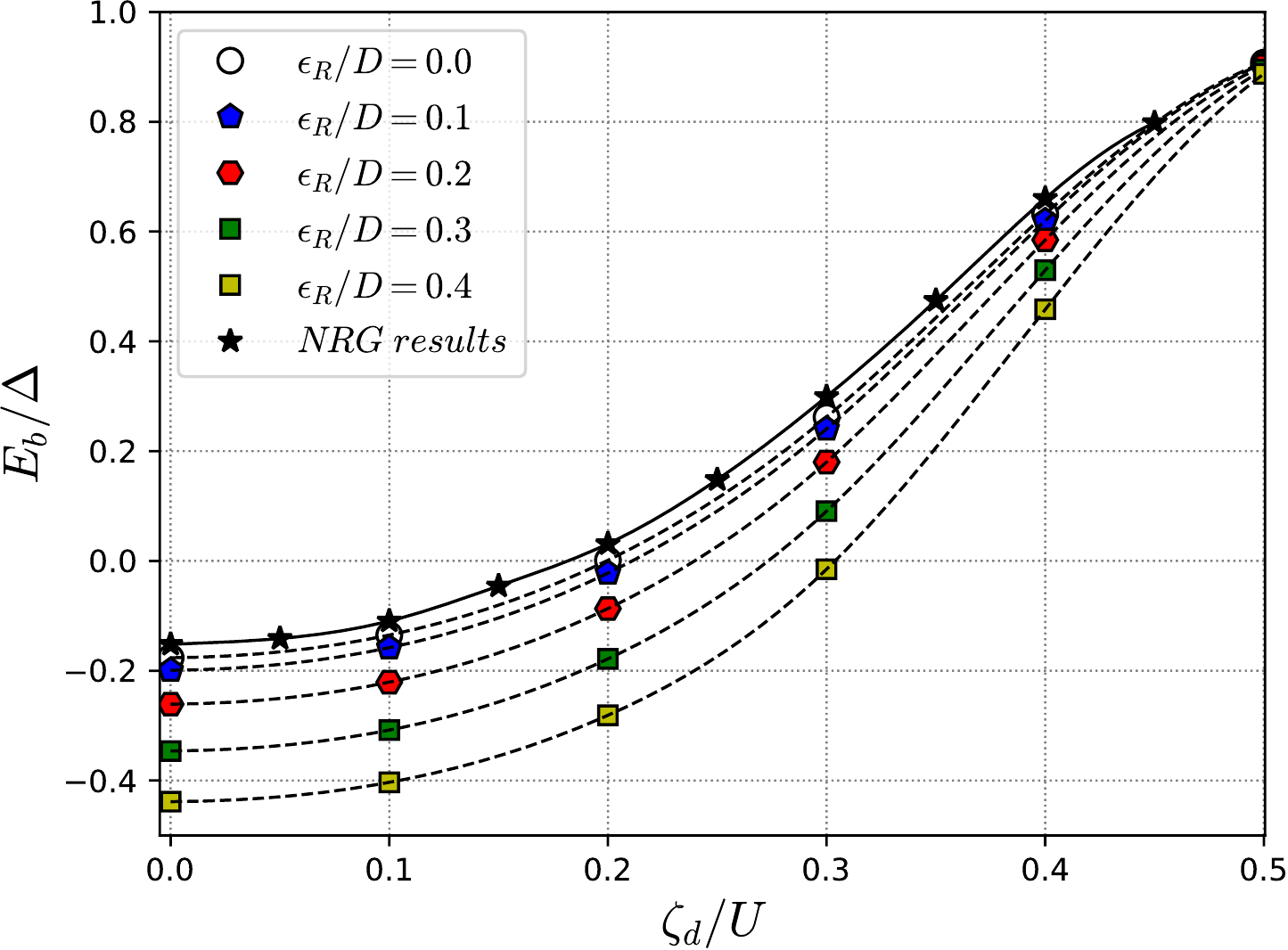}
\caption{(Color online) Energy of one branch of Shiba states for different values of the Rashba SOC coupling, as a function of the assymetry factor $\zeta_d$. NRG results for zero Rashba SOC coupling from Ref. 
\onlinecite{Bauer07_NRG_Anderson_model_in_BCS_superconductor}}
\label{asymmetry}
\end{figure}

Defining the asymmetry parameter as\cite{Bauer07_NRG_Anderson_model_in_BCS_superconductor} 
$\zeta_d=\epsilon_d+U/2$, such that the symmetrical point corresponds to $\zeta_d=0$, in Fig. \ref{asymmetry} we show the effect of the Rasba coupling in the position of the Shiba states when $\zeta_d$ is varied.  For the parameters of the figure, at $\zeta_d=0$ the system is in the doublet phase and in consequence increasing the Rashba SOC coupling at that point drives the system deeper into the doublet phase (the energy of the Shiba states decreases from $0$ as $\alpha_R$ increases). As the system enters the asymmetric regime, the critical value $\zeta_{d,\text{c}}$ at which the transition occurs increases as  $\alpha_R$ increases. Hence we can see that also in the assymetric case the Rashba SOC also favours the doublet phase, i.e., weakens the Kondo regime. 

\subsection{Universality and Kondo Temperature.}
\label{Universality}

In this section, we discuss an important result of our work, the evolution of the Kondo temperature as a function of the Rashba SOC coupling. In addition, we discuss the universality of the model, following the work done previously by Yoshioka and Ohashi \cite{Yoshioka00_NRG_Anderson_impurity_on_SC} for the case $\alpha_R=0$. For $\alpha_R=0$, we recall that the Kondo temperature is defined as\cite{Haldane_1978}: 
\begin{align}
T_K &= 0.364 \sqrt{\frac{2 \Gamma_0 U}{\pi}} \exp\left[{\frac{\epsilon_d \left(\epsilon_d+U\right)\pi }{2 \Gamma_0 
U}}\right],
\label{TK_Haldane}
\end{align}
which for the symmetric case reduces to:
\begin{align}
T_K &= 0.182 U \sqrt{\frac{8 \Gamma_0 }{\pi U}} \exp\left[-{\frac{\pi U }{8 \Gamma_0}}\right].
\label{TK_Haldane_sym} 
\end{align}
With this expression, Yoshioka and Ohashi \cite{Yoshioka00_NRG_Anderson_impurity_on_SC} showed that, within the Kondo regime $\pi \Gamma_0 < U$, the energy of the Shiba state $E_b/\Delta$ is a \emph{universal} function of $T_K/\Delta$. As mentioned before, when $\pi \Gamma_0 > U$ the charge fluctuations in the dot inhibit the formation of a local magnetic moment and the system is away from the Kondo regime. It was also shown that universality breaks down well inside the doublet phase, for values of the SC gap $\Delta$ larger than the critical $\Delta_\text{c}$. 

Since in our single-impurity 1D case the only important effect of the Rasbha coupling is to lower the effective hybridization 
$\Gamma\left(\epsilon_R/D \right)$ [see Eq. (\ref{newGamma})], it can be seen that the universality found for $\alpha_R=0$ 
\cite{Yoshioka00_NRG_Anderson_impurity_on_SC}, will also occur at $\alpha_R \neq 0$ if we define the Kondo temperature with a generalization of Haldane's formula for $\epsilon_R$ finite, as was done in Ref. 
\onlinecite{Wong16_Influence_Rashba_spin-orbit_coupling_Kondo_effect}: 
\begin{figure}[t]
\includegraphics[width=8.5cm]{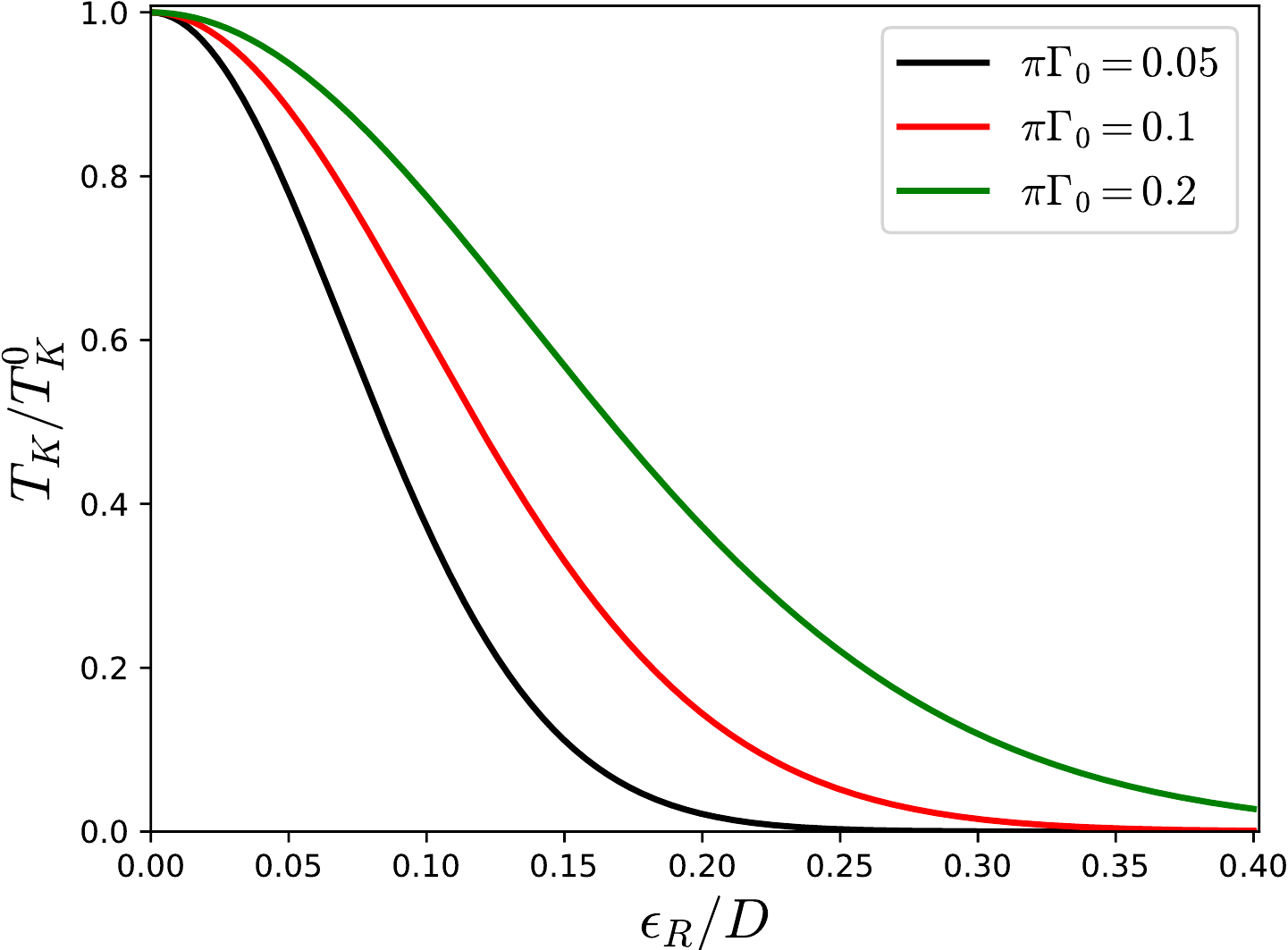}
\caption{(Color online) Normalized Kondo Temperature according to Eq. (\ref{TK_Haldane2}) as a function of the normalized Rashba energy. In this calculation we have used the value $U=0.5D$.}
\label{TK}
\end{figure}
\begin{equation}
 T_K = 0.364 \sqrt{\frac{2 \Gamma\left(\epsilon_R/D\right) U}{\pi}} \exp\left[{\frac{\epsilon_d 
\left(\epsilon_d+U\right)\pi }{2 \Gamma\left(\epsilon_R/D\right) U}}\right]. 
\label{TK_Haldane2} 
\end{equation}
In Fig. \ref{TK} we show $T_K$ given by Eq. (\ref{TK_Haldane2}) as a function of $\epsilon_R/D$. Since the Rashba coupling always lowers the density of states (with respect to $\rho_0$), this  implies that the Kondo temperature is always decreased when $\alpha_R$ (and hence $\epsilon_R$) is increased. 

We stress that the above phenomenology is a consequence of the reduced dimensionality of the 1D nanowire, and that in a 2D system the situation might be different. The effect of the Rashba SOC on $T_K$ in the 2D case has been treated in many previous works \cite{Malecki07_Two_Dimensional_Kondo_Model_with_Rashba_Spin-Orbit_Coupling, Zarea2012_Enhancement_Kondo_Effect_through_Rashba_Spin-Orbit_Interactions,Isaev12_Kondo_effect_in_the_presence_of_spin-orbit_coupling, Wong16_Influence_Rashba_spin-orbit_coupling_Kondo_effect, Yanagisawa12_Kondo_Effect_Spin–Orbit_Coupling, Li18_Rashba-induced_Kondo_screening_magnetic_impurity_two-dimensional}, with very different conclusions (i.e., $T_K$ can either increase, remain constant, or decrease). In particular, in a 2D system the Rashba SOC mixes the spin and orbital momenta of the conduction electrons, and this mixing results in an apparent effective two-channel Anderson or Kondo Hamiltonian. However, as Zitko and Bon{\v c}a  explain\cite{Zitko11_Kondo_effect_in_the_presence_of_Rashba_spin-orbit_interaction}, solving the problem exactly always results in a single-channel model. Therefore, in 2D and near the Fermi level the total density of states does not change with the Rashba SOC, and the Kondo temperature is only weakly affected, linearly increasing or decreasing depending on the impurity parameters \cite{Zitko11_Kondo_effect_in_the_presence_of_Rashba_spin-orbit_interaction}. It is also worth to mention that it has been claimed \cite{Li18_Rashba-induced_Kondo_screening_magnetic_impurity_two-dimensional} that the mixing of spin and orbital momentum of the conduction electrons  leads to a Rashba-dependent effective SC gap $\Delta\rightarrow \Delta\left(\epsilon_R\right)$, something that does not occur in 1D case. On the other hand, in the purely 1D case the influence of the Rashba SOC on the Kondo temperature has also been studied using a Schrieffer-Wolff transformation and a ``poor's man'' scaling approach, and it was shown that the coupling $J$ of the resulting  Kondo model \emph{increased} with the Rashba SOC\cite{deSousa16_Kondo_effect_quantum_wire_with_spin-orbit_coupling} . Nevertheless, since $T_K$ in the Kondo model essentially depends on the product $\rho_0 J$, and taking into account the renormalization of $\rho_0$ in Eq. (\ref{eq:rho0R}), it is clear that the increase in $J$ must be overcome by the decrease in $\rho_0$, in such a way that the overall effect is a \emph{net decrease} of the product  $\rho_0 J$ with the Rashba SOC.

\begin{figure}[t]
\includegraphics[width=8.5cm]{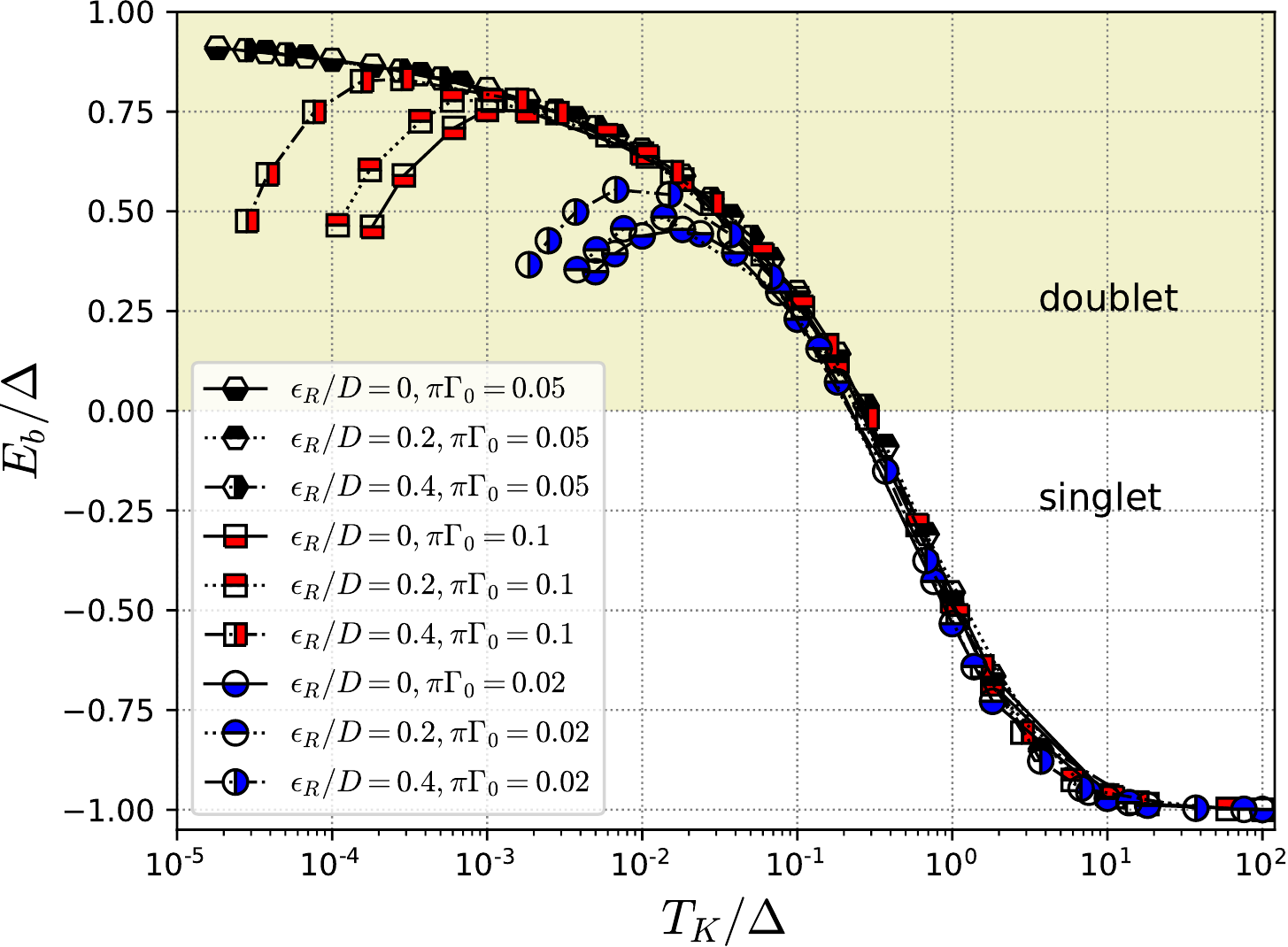}
\caption{(Color online) Universality for the particle-hole symmetric case. Energy of the Shiba states, in units of the SC gap, for different values of $\epsilon_R/D$ as a function of $T_K/\Delta$. $U=0.5$}
\label{universality}
\end{figure}

\begin{figure}[t]
\includegraphics[width=8.5cm]{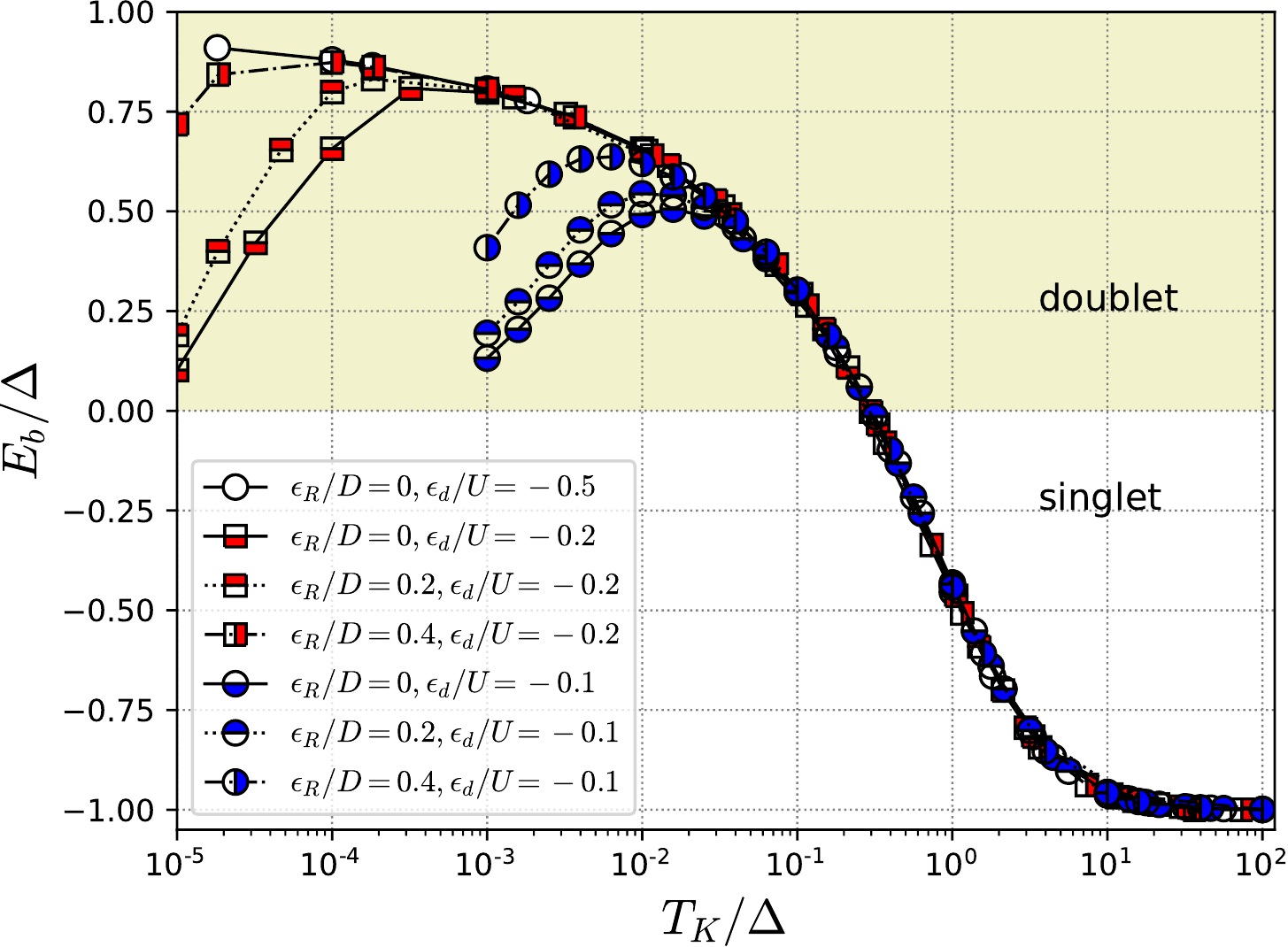}
\caption{(Color online) Universality away from the particle-hole symmetric case, for different values of $\epsilon_R/D$. In 
this figure $\pi \Gamma_0=0.05$ and $U=0.5$.}
\label{universality2}
\end{figure}

Finally, we analize the the universality of the model, following the work of Yoshioka and Ohashi \cite{Yoshioka00_NRG_Anderson_impurity_on_SC} for the case without Rashba SOC. In Figs. \ref{universality} and \ref{universality2} we plot the energy of the Shiba states as a function of $T_K/\Delta$, with $T_K$ defined as in Eq. (\ref{TK_Haldane2}). Here we only vary the SC gap $\Delta$, with $U$, $\epsilon_d$ and $\Gamma_0$ fixed \cite{note_universality}. In Fig. \ref{universality} we show the DMRG results for the symmetric case $\epsilon_d=-U/2$ and in Fig. \ref{universality2} the results away from the particle-hole symmetric case, with fixed $\pi \Gamma_0$. The results corresponding to $\alpha_R=0$ match very well those obtained with NRG by Yoshioka and Ohashi \cite{Yoshioka00_NRG_Anderson_impurity_on_SC} (not shown to maintain the clarity of the figure). In the Kondo singlet phase (lower half region of the figures) the results are universal independently of the value of $\alpha_R$. This is expected since	, as was shown in Eq. (\ref{newGamma}), the effect of the Rashba coupling can be reduced to a change in the hybridization, but again we note that in 2D the results are different due to the renormalization of the SC gap  $\Delta$\cite{Li18_Rashba-induced_Kondo_screening_magnetic_impurity_two-dimensional}. Therefore, when the generalized Kondo temperature Eq. (\ref{TK_Haldane2}) is used, all the curves collapse into a single one, and hence we can argue that the 0-$\pi$ transition occurs at the universal value $T_K/\Delta\simeq 0.3$, even in the presence of Rasbha SOC, at least in our 1D case. In the doublet phase where  $T_K/\Delta\ll 1$, universality is lost (see down turn of the curves) and the value of $\Delta$ needed to achieve this non-universal regime  is larger as $\alpha_R$ increases.

\section{Summary and perspectives}
\label{conclusions}

We have studied the effect of the Rashba spin-orbit coupling present in a one-dimensional superconducting nanowire coupled to a single-level quantum dot, and have analyzed its influence on the 0-$\pi$ transition and  Kondo temperature. Our work was motivated mainly by recent experimental systems where the Rashba spin-orbit coupling has been identified as an unavoidable ingredient, such as semiconductor nanowires proximitized by nearby bulk superconductors or the surface of superconducting materials (such as Al or Pb) with large atomic numbers and strong intra-atomic spin-orbit interaction. Those systems have been used as experimental platforms with magnetic impurities or quantum dots where Kondo and Shiba physics has been revealed. 

We have modeled the quantum dot by means of the Anderson impurity model. In order to solve the many-body problem, we have implemented the DMRG, in combination with logarithmic discretization of the conduction band and a subsequent mapping onto a Wilson chain Hamiltonian, in order to accomodate subgap Shiba states with exceedingly long localization lengths. We have benchmarked  and tested this method against previous results obtained with the NRG techninque in the absence of  Rashba SOC, with excellent agreement.

We have particularly studied the 0-$\pi$ singlet-to-doublet phase transition and the position of the subgap (Shiba) states, showing in detail their dependence on the Rashba coupling. By the means of a  straightforward unitary transformation, we have been able to show that in a 1D geometry, the most important effect of the Rashba coupling can be accounted for in a reduction in the density of normal states in the conduction band. Using this result and a generalized Haldane's formula for $T_K$ we have shown that the Kondo temperature it is always lowered by the Rashba coupling in this one dimensional case. Physically, this has the indirect effect of favoring the doublet phase.

The excellent results obtained with DMRG open the possibility of studying chains or clusters of impurities coupled to normal or topological superconductors. This is an interesting perspective since this kind of systems have only be studied analytically in non-interacting systems where the Kondo effect is absent. 
 
\section*{Acknowledgments}

We thank Luis O. Manuel for very useful discussions. This work was partially supported by CONICET (Grants PIP 112-20150100-364 and PIP 1122015010036) and ANPCyT (PICT 2017-2081), Argentina.

\bibliographystyle{apsrev}


\begin{thebibliography}{76}
\bibitem[{not({\natexlab{a}})}]{note1}
\bibinfo{note}{It is worth mentioning that if we had departed from a
  tight-binding model, the result of eq. \ref{eq:rho0R} would still be valid}.
\bibitem[{not({\natexlab{b}})}]{note_universality}
\bibinfo{note}{Of course when $E_b >0$ the GS is a doublet and the system is no
  longer in the Kondo regime. Nevertheless we still plot the energy of the
  Shiba states against $T_K$ given by Eq. (\ref{TK_Haldane2})}.
\end{thebibliography}

\end{document}